\documentclass[aps,prl,reprint,groupedaddress]{revtex4-1}
\usepackage{graphicx}
\usepackage{amsmath,amssymb}
\usepackage{amsfonts}
\usepackage{mathtools}
\usepackage[utf8]{inputenc}
\usepackage[colorlinks=true,linkcolor=blue,citecolor=blue,filecolor=cyan,urlcolor=blue,breaklinks=true]{hyperref}
\usepackage{subfigure}
\usepackage{psfrag}
\usepackage{hyperref}
\usepackage{color}
\usepackage{microtype}
\DisableLigatures[f]{encoding = *, family = *}
\newcommand{\diag}{\mathrm{diag}}

\definecolor{blue}{RGB}{0,50,200}
\definecolor{red}{RGB}{170,33,44}

\begin{document}
\title{Parameter inference and nonequilibrium identification for Markovian systems \\based on coarse-grained observations}
\author{Bingjie Wu$^{1}$}
\author{Chen Jia$^1$}\email{Correspondence: chenjia@csrc.ac.cn}
\affiliation{$^1$ Applied and Computational Mathematics Division, Beijing Computational Science Research Center, Beijing 100193, China.}

\parskip 1mm
\def\F{\mathcal{A}}
\def\Q{\mathcal{Q}}

\begin{abstract}
Most experiments can only detect a set of coarse-grained clusters of a molecular system, while the internal microstates are often inaccessible. Here, based on an infinitely long coarse-grained trajectory, we obtain a set of sufficient statistics which extracts all statistic information of coarse-grained observations. Based on these sufficient statistics, we set up a theoretical framework of parameter inference and nonequilibrium identification for a general Markovian system with an arbitrary number of microstates and arbitrary coarse-grained partitioning. Our framework can identify whether the sufficient statistics are enough for empirical estimation of all unknown parameters and we can also provide a quantitative criterion that reveals nonequilibrium. Our nonequilibrium criterion generalizes the one obtained [J. Chem. Phys. 132:041102 (2010)] for a three-state system with two coarse-grained clusters, and is capable of detecting a larger nonequilibrium region compared to the classical criterion based on autocorrelation functions.
\end{abstract}


\maketitle

\emph{Introduction} --- Mesoscopic molecular systems are widely modeled as a Markov process with a large number of microstates \cite{cornish2013fundamentals, sakmann2013single, golding2005real}.  In experiments, it often occurs that only a set of coarse-grained clusters can be detected, while the internal microstates of the system are often inaccessible or indistinguishable. For instance, using live-cell imaging, one can obtain the time trace of the copy number of a protein in a single cell; however, it is difficult to determine whether the gene is in an active or an inactive state \cite{golding2005real}. Due to the inability to accurately identify all microstates, the data obtained is usually the time trace of some coarse-grained states, which only retain a small degrees of freedom of the system \cite{seifert2019stochastic, esposito2012stochastic, teza2020exact}. Other well-known examples of partial observations include ion channel opening \cite{sakmann2013single}, molecular docking and undocking to a sensor \cite{skoge2013chemical}, and flagellar motor switches \cite{berg2003rotary}.

Given a sufficiently long trajectory of coarse-grained states, two natural and crucial questions arise: (i) is it possible to determine the transition topology and even all transition rates between all microstates? (ii) If complete recovery of transition rates is impossible, is it still possible to determine whether the system is in an equilibrium state or in a nonequilibrium steady state (NESS) and even estimate the values of some macroscopic thermodynamic quantities? The detection of nonequilibrium is important since in an NESS, there are nonzero net fluxes between microstates, indicating that the system is externally driven with concomitant entropy production \cite{qian2000pumped, li2002kinetic, witkoskie2006testing, tu2008nonequilibrium, amann2010communications, jia2015second, skinner2021estimating, harunari2022learn, van2022thermodynamic, van2023thermodynamic, ghosal2023entropy}. Over the past two decades, numerous studies have partially answered these questions; however, there is still a lack of a unified theory for general systems.

Among these studies, some \cite{bruno2005using, flomenbom2005can, flomenbom2006utilizing, flomenbom2008universal} focus on transition topology inference; some \cite{deng2003identifying, xiang2006identifying, xiang2024identifying} focus on transition rate inference with a given transition topology; some \cite{qian2000pumped, li2002kinetic, witkoskie2006testing, tu2008nonequilibrium, amann2010communications, jia2015second} investigate nonequilibrium detection based on a two-state coarse-grained trajectory. Recently, several studies have estimated the values of some macroscopic thermodynamic quantities, such as entropy production and cycle affinities, based on partial observations \cite{skinner2021estimating, harunari2022learn, van2022thermodynamic, van2023thermodynamic, ghosal2023entropy}. In this study, we develop a sufficient statistics approach and set up a theoretical framework of parameter recovery and nonequilibrium detection for a general Markovian system with a given transition topology. Our results only depend on data available from a sufficient long trajectory between coarse-grained states.

\begin{figure}[htb!]
\centering\includegraphics[width = 0.48\textwidth]{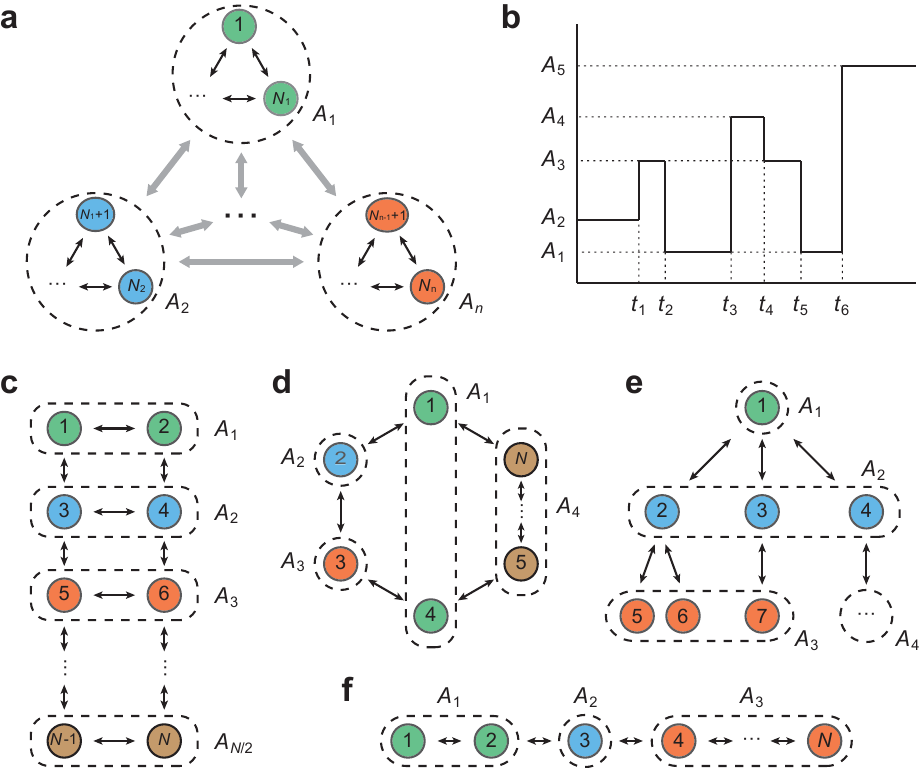}
\caption{\textbf{Model}.
\textbf{(a)} An $N$-state Markovian system with $n$ coarse-grained states $A_1,\cdots,A_n$, each composed of multiple microstates. \textbf{(b)} Illustration of an infinitely long trajectory of coarse-grained states.
\textbf{(c)} The ladder model.
\textbf{(d)} The cyclic model.
\textbf{(e)} The tree model.
\textbf{(f)} The linear model. }\label{model}
\end{figure}

\emph{Model} --- We consider an $N$-state molecular system modeled by a continuous-time Markov chain $(X_s)_{s\ge 0}$ with microstates $x = 1,\cdots,N$ and generator matrix $Q = (q_{xy})$, where $q_{xy}$ denotes the transition rate from state $x$ to state $y$ whenever $x\neq y$ and $q_{xx} = -\sum_{y\neq x}q_{xy}$. Recall that the transition diagram of the Markovian system is a directed graph with vertex set $V = \{1,\cdots,N\}$ and edge set $E = \{(x,y):q_{xy}>0\}$. Experimentally, it often occurs that not all microstates can be observed --- what can be observed are often some coarse-grained states, each being composed of multiple microstates. Specifically, we assume that the $N$ microstates can be divided into $n$ coarse-grained states $A_1,\cdots,A_n$ (Fig. \ref{model}(a)). We also assume that the observation is an infinitely long trajectory of coarse-grained states (Fig. \ref{model}(b)) \cite{flomenbom2005can,  flomenbom2006utilizing, flomenbom2008universal}.

Based on the coarse-grained clusters, the generator matrix $Q$ can be represented as the block form
\begin{equation}\label{fm:Q}
Q = \begin{pmatrix}
Q_{A_1} & Q_{A_1A_2} & \cdots & Q_{A_1A_n}\\
Q_{A_2A_1} & Q_{A_2} & \cdots & Q_{A_2A_n}\\
\vdots & \vdots & \ddots & \vdots\\
Q_{A_nA_1} & Q_{A_nA_2} & \cdots & Q_{A_n}
\end{pmatrix}.
\end{equation}
The steady-state distribution $\mathbf{\pi} = (\pi_{A_1},\cdots,\pi_{A_n})$ of the system must satisfy $\mathbf{\pi}Q = \mathbf{0}$, where $\mathbf{0} = (0,\cdots,0)$ is the zero vector. For simplicity, we assume that each diagonal block $Q_{A_i}$ has different eigenvalues $-\lambda^i_{1},\cdots,-\lambda^i_{|A_i|}$ (any matrix with repeated eigenvalues can be approximated by matrices with different eigenvalues to any degree of accuracy \cite{horn2012matrix}). Then there exists an invertible matrix $\Phi_{A_i}$ such that $Q_{A_i}=\Phi^{-1}_{A_i}\Lambda_{A_i}\Phi_{A_i}$, where $\Lambda_{A_i}=\diag(-\lambda^i_{1},\cdots,-\lambda^i_{|A_i|})$ is a diagonal matrix. Note that each row of $\Phi_{A_i}$ is an eigenvector of $Q_{A_i}$. For convenience, the sum of elements of each eigenvector is normalized to one, i.e, $\Phi_{A_i}\mathbf{1}^T = \mathbf{1}^T$, where $\mathbf{1}=(1,\cdots,1)$. Moreover, we set
\begin{equation}\label{definition}
\begin{split}
\Lambda_{A_iA_j} &= \Phi_{A_i}Q_{A_iA_j}\Phi_{A_j}^{-1} = (\lambda^{ij}_{kl}),\\
\alpha_{A_i} &= \pi_{A_i}\Phi_{A_i}^{-1}=(\alpha^i_{k}).
\end{split}
\end{equation}
These quantities will play a crucial role in our analysis.

\emph{Sufficient statistics} ---  In what follows, we assume that the system has reached a steady state. We next examine which information can be extracted from the infinitely long coarse-grained trajectory (Fig. \ref{model}(b)). Clearly, what can be observed from the trajectory are the jump times and the coarse-grained states before and after each jump within any finite period of time. In other words, for any given jump times $0\leq t_1<...<t_m\leq t$ and any coarse-grained states $A_{i_1},\cdots,A_{i_{m+1}}$, we can estimate the following probability (density) since the trajectory is infinitely long \cite{fredkin1986aggregated}:
\begin{equation}\label{fm:ob}
\begin{split}
\mathbb{P}(X_s\in A_{i_1}(s< t_1),\cdots,X_s\in A_{i_{m+1}}(t_m\le s\le t)).
\end{split}
\end{equation}
These probabilities are actually all statistical information that can be extracted from coarse-grained observations.

We first consider the case where there is no jump before time $t$. Based on the notation in Eq. \eqref{definition}, it is clear that
\begin{equation}\label{fm:ob1}
\begin{split}
&\mathbb{P}\left(X_s\in A_i(0\le s\le t)\right) = \pi_{A_i}e^{Q_{A_i}t}\mathbf{1}^T\\
&=\alpha_{A_i}e^{\Lambda_{A_i}t}\mathbf{1}^T=\sum_{k}\alpha^i_{k} e^{-\lambda^i_{k}t}.
\end{split}
\end{equation}
Based on coarse-grained observations, we can estimate the probability on the left-hand side of the above equation for each time $t$. Since time $t$ is arbitrary, we can estimate the values of all $\alpha^i_k$ and $\lambda^i_k$, and hence $\alpha_{A_i}$ and $\Lambda_{A_i}$ can be determined. Similarly, if there is only one jump before time $t$, then for any $i\neq j$, we have
\begin{equation}\label{fm:ob2}
\begin{split}
&\mathbb{P}\left(X_s\in A_{i}( 0\le s< t_1),X_s\in A_{j}( t_1\le s\le t)\right)\\
&=\pi_{A_i}e^{Q_{A_i}t_1}Q_{A_iA_j}e^{Q_{A_j}(t-t_1)}\mathbf{1}^T\\
&=\alpha_{A_i}e^{\Lambda_{A_i}t_1}\Lambda_{A_iA_j}e^{\Lambda_{A_j}(t-t_1)}\mathbf{1}^T.
\end{split}
\end{equation}
Since times $t_1$ and $t$ are arbitrary and since we have determined $\alpha_{A_i}$ and $\Lambda_{A_i}$, we can also determine $\Lambda_{A_iA_j}$ from coarse-grained observations for any $i\neq j$ \cite{supp}.

Similarly to Eqs. \eqref{fm:ob1} and \eqref{fm:ob2}, it can be proved that \cite{supp}
\begin{eqnarray}\label{likelihood}
&\mathbb{P}(X_s\in A_{i_1}(0\le s< t_1),\cdots, X_s\in A_{i_{m+1}}(t_m\le s\le t)) \nonumber\\
&=\pi_{A_{i_1}}e^{Q_{A_{i_1}}t_1}\cdots Q_{A_{i_m}A_{i_{m+1}}}e^{Q_{A_{i_{m+1}}}(t-t_m)}\mathbf{1}^T\\
&=\alpha_{A_{i_1}}e^{\Lambda_{A_{i_1}}t_1}\cdots\Lambda_{A_{i_m}A_{i_{m+1}}}e^{\Lambda_{A_{i_{m+1}}}(t-t_m)}\mathbf{1}^T. \nonumber
\end{eqnarray}
This shows that the probability given in Eq. \eqref{fm:ob} can be represented by $\alpha_{A_i}$, $\Lambda_{A_i}$, and $\Lambda_{A_iA_j}$. Hence these three quantities contain all coarse-grained statistical information. In fact, $\alpha_{A_i}$, $\Lambda_{A_i}$, and $\Lambda_{A_iA_j}$ are not independent. Since $Q\mathbf{1}^T = \mathbf{0}^T$, it is clear that $\Lambda\mathbf{1}^T = \mathbf{0}^T$, where
\begin{equation}
\Lambda = \begin{pmatrix}
\Lambda_{A_1} & \Lambda_{A_1A_2} & \cdots & \Lambda_{A_1A_n}\\
\vdots & \vdots & \ddots & \vdots\\
\Lambda_{A_nA_1} & \Lambda_{A_nA_2} & \cdots & \Lambda_{A_n}
\end{pmatrix}.
\end{equation}
This implies that $\lambda^i_k=\sum_{j,l}\lambda^{ij}_{kl}$, where $\lambda^{ij}_{kl}$ are the elements of $\Lambda_{A_iA_j}$ defined in Eq. \eqref{definition}. Hence $\Lambda_{A_i}$ can be determined by all $\Lambda_{A_iA_j}$. Since $\pi Q = \mathbf{0}$, we have $\alpha\Lambda = \mathbf{0}$, where $\alpha=(\alpha_{A_1},\cdots,\alpha_{A_n})$. This equation, together with the normalization condition
\begin{equation}
\alpha\mathbf{1}^T = \sum_{i=1}^n\pi_{A_i}\Phi_{A_i}^{-1}\mathbf{1}^T = \sum_{i=1}^n\pi_{A_i}\mathbf{1}^T = 1,
\end{equation}
shows that $\alpha_{A_i}$ can also be determined by all $\Lambda_{A_iA_j}$.

Recall that a family of statistics that contain all statistical information of the observations are called sufficient statistics \cite{fisher1922mathematical}. Since the probability in Eq. \eqref{fm:ob} can be represented by $\alpha_{A_i}$, $\Lambda_{A_i}$, and $\Lambda_{A_iA_j}$ and since $\alpha_{A_i}$ and $\Lambda_{A_i}$ can be determined by all $\Lambda_{A_iA_j}$, it is clear that all $\Lambda_{A_iA_j}$ are the sufficient statistics for infinitely long coarse-grained observations. In particular, the number of these sufficient statistics is given by
\begin{equation}
S = \sum_{1\le i\neq j\le n}|A_i||A_j| = 2\sum_{1\le i<j\le n}|A_i||A_j|.
\end{equation}
Note that based on coarse-grained observations, two systems with different generator matrices but having the same $\Lambda_{A_iA_j}$ are statistically indistinguishable.

\emph{Parameter inference} --- In practice, a crucial question is whether all transition rates of a Markovian system with a given transition topology can be inferred from an infinitely long coarse-grained trajectory. Note that each transition rate $q_{xy}$ of the system corresponds to a direct edge in the transition diagram $(V,E)$. Hence the system has $|E|$ unknown parameters, where $|E|$ denotes the number of elements in the edge set $E$. Note that each $\Lambda_{A_iA_j}$ is uniquely determined by all transition rates and in the following, we rewrite it as $\Lambda_{A_iA_j}(q_{xy})$. Based on coarse-grained observations, we can obtain estimates of the sufficient statistics $\Lambda_{A_iA_j}$ and hence all transition rates should satisfy the following set of equations:
\begin{equation}\label{fm:pa1}
\Lambda_{A_iA_j}(q_{xy}) = \tilde{\Lambda}_{A_iA_j},\;\;\;\forall\;1\le i\neq j\le n,
\end{equation}
where $\tilde{\Lambda}_{A_iA_j}$ are the estimates of $\Lambda_{A_iA_j}$. Clearly, Eq. \eqref{fm:pa1} has $|E|$ unknown parameters and $S$ equations, where $S$ is the number of sufficient statistics. Therefore, we obtain the following criteria regarding parameter inference: (i) when $S<|E|$, it is impossible to determine all transition rates of the system; (ii) when $S\ge |E|$, all transition rates can be determined if and only if Eq. \eqref{fm:pa1} has a unique solution. In general, it is very difficult to give a simple criterion for the unique solvability of Eq. \eqref{fm:pa1}; however, it can be checked numerically using, e.g., Gröbner basis computation \cite{weispfenning1992comprehensive} and can be proved theoretically in some simple examples. Next we focus on three examples.

1) The ladder model (Fig. \ref{model}(c)). This model is widely used to model allostery of receptors in living cells with strong cooperativity and high sensitivity \cite{monod1965nature}. Consider a receptor with two conformational states and $n$ ligand binding sites. According to the conformational state and the number of occupied binding sites, each receptor can be modeled by a Markovian system with $N = 2n$ microstates. Due to technical limitations, we are often unable to distinguish between the two conformational states. The $N$ microstates and $n$ coarse-grained states are shown in Fig. \ref{model}(c). Clearly, we have $|A_1|=\cdots=|A_n|=2$ and there are $|E|=3N-4$ unknown parameters for the system. Note that for any $n\ge 2$, we have
\begin{equation}
S = N^2-2N \ge 3N-4 = |E|.
\end{equation}
Hence all parameters of the ladder model can be inferred if and only if Eq. \eqref{fm:pa1} has unique solution. In \cite{supp}, we show that Eq. \eqref{fm:pa1} is indeed uniquely solvable for the ladder model.

2) The cyclic model (Fig. \ref{model}(d)). Many crucial cellular biochemical processes can be modeled as cyclic Markovian systems such as conformational changes of enzymes and ion channels \cite{cornish2013fundamentals, sakmann2013single}, phosphorylation-dephosphorylation cycle \cite{qian2007phosphorylation}, cell cycle progression \cite{jia2021frequency}, and gene state switching \cite{jia2023analytical}. An $N$-state cyclic model has $|E| = 2N$ unknown parameters. If there are only two coarse-grained states ($n = 2$), then it is impossible to infer all unknown parameters in the case of $|A_1| = 1$ and $|A_2| = N-1$ since $S = 2(N-1)<|E|$. For any other coarse-grained partitioning, we have $S\geq 4(N-2)\geq |E|$ and hence parameter inference is generally possible. When $n\geq 3$, we also have $S\ge 4N-6\ge|E|$, and hence a complete parameter recovery can be made if Eq. \eqref{fm:pa1} can be uniquely solved. In \cite{supp}, we show that Eq. \eqref{fm:pa1} is indeed uniquely solvable for the four-state cyclic model under any coarse-grained partitioning whenever $S\geq |E|$.

3) The tree and linear models (Fig. \ref{model}(e)). We finally focus on an $N$-state system whose transition diagram is a tree. For any coarse-grained partitioning, it is easy to check that $S\geq 2(N-1)=|E|$. Hence all transition rates can be inferred if Eq. \eqref{fm:pa1} has unique solution. In particular, a system with linear transitions (Fig. \ref{model}(f)) can be viewed as a tree. The linear model also widely used in biochemical studies \cite{ullah2012simplification}. In \cite{supp}, we show that Eq. \eqref{fm:pa1} is indeed uniquely solvable for the linear model if there are only two coarse-grained states ($n = 2$) with $|A_1| = 1$ and $|A_2| = N-1$.

\emph{Nonequilibrium identification} --- When the number of sufficient statistics is less than the number of unknown parameters, it is impossible to determine all transition rates from coarse-grained observations. However, we may still identify whether the system is in an NESS. Our Ness criterion is based on the sufficient statistics $\Lambda_{A_iA_j}$.
Recall that once $\Lambda_{A_iA_j}$ are determined, $\alpha_{A_i}$ are automatically determined by solving $\alpha\Lambda = \mathbf{0}$ and $\alpha\mathbf{1}^T = 1$. In \cite{supp}, we prove that if the system is in equilibrium, then the entries of $\alpha_{A_i}$ and $\Lambda_{A_iA_j}$ are all real numbers, and the following two conditions must be satisfied:\\
(i) (coarse-grained probability distribution condition)
\begin{equation}\label{fm:criterion1}
\alpha^i_k\geq 0.
\end{equation}
(ii) (coarse-grained detailed balance condition)
\begin{equation}\label{fm:criterion2}
\alpha^i_{k}\lambda^{ij}_{kl}= \alpha^j_{l} \lambda^{ji}_{lk}.
\end{equation}
Recall that $\alpha \mathbf{1}^T =1$. When the system is in equilibrium, we have $\alpha^i_k\geq 0$ and thus $\alpha = (\alpha_{A_1},\cdots,\alpha_{A_n})$ is a probability distribution. This is why Eq. \eqref{fm:criterion1} is called the coarse-grained probability distribution condition. On the other hand, in equilibrium, the system satisfies the detailed balance condition $\pi_xq_{xy}=\pi_yq_{yx}$. Eq. \eqref{fm:criterion2} can be viewed as the coarse-grained version of the detailed balance condition. The above result implies that if any one of Eqs. \eqref{fm:criterion1} and \eqref{fm:criterion2} is violated, then the system must be in an NESS. This gives a general criterion for detecting nonequilibrium based on coarse-grained observations.

For a three-state cyclic system with two coarse-grained states $A_1=\{1\}$ and $A_2=\{2,3\}$, we have seen that it is impossible to infer all parameters. In \cite{amann2010communications}, the authors showed that this system is in an NESS when
\begin{equation}\label{fm:criterion3}
TL^2 > LSM-M^2,
\end{equation}
where $L = q_{12}+q_{13}$, $S=q_{21}+q_{23}+q_{31}+q_{32}$, $T = q_{21}q_{31}+q_{21}q_{32}+q_{23}q_{31}$, and $M = q_{12}(q_{23}+q_{31}+q_{32})+q_{13}(q_{21}+q_{23}+q_{32})$ are another set of sufficient statistics for the three-state system that can be determined by coarse-grained observations \cite{amann2010communications, jia2015second}. Our result can be viewed as an extension of Eq. \eqref{fm:criterion3} to complex molecular systems with an arbitrary number of microstates and an arbitrary number of coarse-grained states.

In particular, when $N = 3$ and $n = 2$, our criterion reduces to Eq. \eqref{fm:criterion3}. This can be seen as follows. First, since $\alpha\Lambda = \mathbf{0}$ and $\Lambda\mathbf{1}^T = \mathbf{0}^T$, it is easy to see that
\begin{equation}\label{identities}
\alpha^1_1\lambda^{12}_{12}=\alpha^2_2\lambda^2_1=\alpha^2_2\lambda^{21}_{21},\;\;\; \alpha^1_1\lambda^{12}_{13}=\alpha^2_3\lambda^2_2=\alpha^2_3\lambda^{21}_{31}.
\end{equation}
Hence the coarse-grained detailed balance condition is satisfied. On the other hand, since $\alpha\mathbf{1}^T = \alpha^1_1+\alpha^2_2+\alpha^2_3 = 1$, it follows from Eq. \eqref{identities} that
\begin{equation}\label{fm:alpha}
\begin{split}
\alpha^1_1 &= \frac{\lambda^{21}_{21}\lambda^{21}_{31}}{\lambda^{21}_{21}\lambda^{21}_{31}
+\lambda^{12}_{12}\lambda^{21}_{31}+\lambda^{12}_{13}\lambda^{21}_{21}},\\
\alpha^2_2 &= \frac{\lambda^{12}_{12}\lambda^{21}_{31}}{\lambda^{21}_{21}\lambda^{21}_{31}
+\lambda^{12}_{12}\lambda^{21}_{31}+\lambda^{12}_{13}\lambda^{21}_{21}},\\
\alpha^2_3 &= \frac{\lambda^{12}_{13}\lambda^{21}_{21}}{\lambda^{21}_{21}\lambda^{21}_{31}
+\lambda^{12}_{12}\lambda^{21}_{31}+\lambda^{12}_{13}\lambda^{21}_{21}}.
\end{split}
\end{equation}
Furthermore, it is easy to check that \cite{supp}
\begin{equation}\label{fm:LSTM}
\begin{split}
&\;\;\lambda^{21}_{21}\lambda^{21}_{31} = T,\;\;\;\lambda^{12}_{12}\lambda^{21}_{31}+\lambda^{12}_{13}\lambda^{21}_{21} = M,\\
&\lambda^{12}_{12}\lambda^{21}_{31}\lambda^{12}_{13}\lambda^{21}_{21} = \frac{T(LSM-M^2-TL^2)}{(\lambda^{12}_{13}-\lambda^{12}_{12})^2}.
\end{split}
\end{equation}
Combining Eqs. \eqref{fm:alpha} and \eqref{fm:LSTM}, we immediately obtain
\begin{equation}
\begin{split}
&\alpha^1_1 = \frac{T}{T+M}\geq 0,\;\;\;\alpha^2_2+\alpha^2_3 = \frac{M}{T+M}\geq 0,\\
&\quad\quad\alpha^2_2\alpha^2_3 = \frac{T(LSM-M^2-TL^2)}{(T+M)^2(\lambda^{12}_{13}-\lambda^{12}_{12})^2}.
\end{split}
\end{equation}
Since $\alpha^1_1\geq0$ and $\alpha^2_2+\alpha^2_3\geq0$, the coarse-grained probability distribution condition is violated if and only if $\alpha^2_2\alpha^2_3 < 0$, which is obviously equivalent to Eq. \eqref{fm:criterion3}.

We now compare our NESS criterion with the classical criterion based on autocorrelation functions. For any observable $\phi$, recall that the autocorrelation function of the system, $B_{\phi}(t) = \mathrm{Cov}(\phi(X_0),\phi(X_t))$, is defined as the steady-state covariance between $\phi(X_0)$ and $\phi(X_t)$. For coarse-grained observations, the observable $\phi$ is usually chosen as $\phi(x) = i$ for all $x\in A_i$. It is well-known \cite{qian2003fundamental} that if the system is in equilibrium, then
\begin{equation}\label{fm:criterion4}
B_{\phi}(t) = \sum_{m=1}^{N-1}c_me^{-\ell_m t},\;\;\;\ell_m>0,\;c_m\geq 0,
\end{equation}
where $-\ell_m<0$ are all nonzero eigenvalues of the generator matrix $Q$ and $c_m\ge 0$ are the coefficients (note that in the case, the autocorrelation function must be monotonic). Hence if there exists some $1\leq m\leq N-1$ such that any one of $\ell_m$ and $c_m$ is negative or not real, then the system must be in an NESS. Clearly, this NESS criterion is stronger than the criterion based on oscillatory or non-monotonic autocorrelation functions \cite{qian2000pumped}.

In \cite{supp}, we have proved a stronger result --- if Eqs. \eqref{fm:criterion1} and \eqref{fm:criterion2} hold, then we also have $\ell_m>0$ and $c_m\geq 0$. This suggests that all NESS scenarios that can identified by the autocorrelation criterion can definitely be identified by our criterion; in other words, our NESS criterion is mathematically stronger than the autocorrelation criterion. In particular, if there are only two coarse-grained states ($n = 2$) with $|A_1| = 1$ and $|A_2| = N-1$, then the two criteria are equivalent; in this case, the number of sufficient statistics is $S = 2(N-1)$ and $\{\ell_1,\cdots,\ell_{N-1},c_1,\cdots,c_{N-1}\}$ is exactly a set of sufficient statistics \cite{supp}. For other coarse-grained partitionings, the two criteria are in general not equivalent and our criterion may extend the NESS region significantly beyond the one identified by the autocorrelation criterion.

We stress that our NESS criterion is only a sufficient condition; there may be some NESS scenarios that fail to be detected by our criterion. However, we prove in the End Matter that for any three-state system with two coarse-grained states ($N = 3$ and $n = 2$), our criterion is also a necessary condition. In other words, if Eqs. \eqref{fm:criterion1} and \eqref{fm:criterion2} are both satisfied, then among all three-state systems having the same sufficient statistics $\Lambda_{A_iA_j}$, there must exist a system which is in equilibrium.

\begin{figure}[htb!]
\centering\includegraphics[width = 0.43\textwidth]{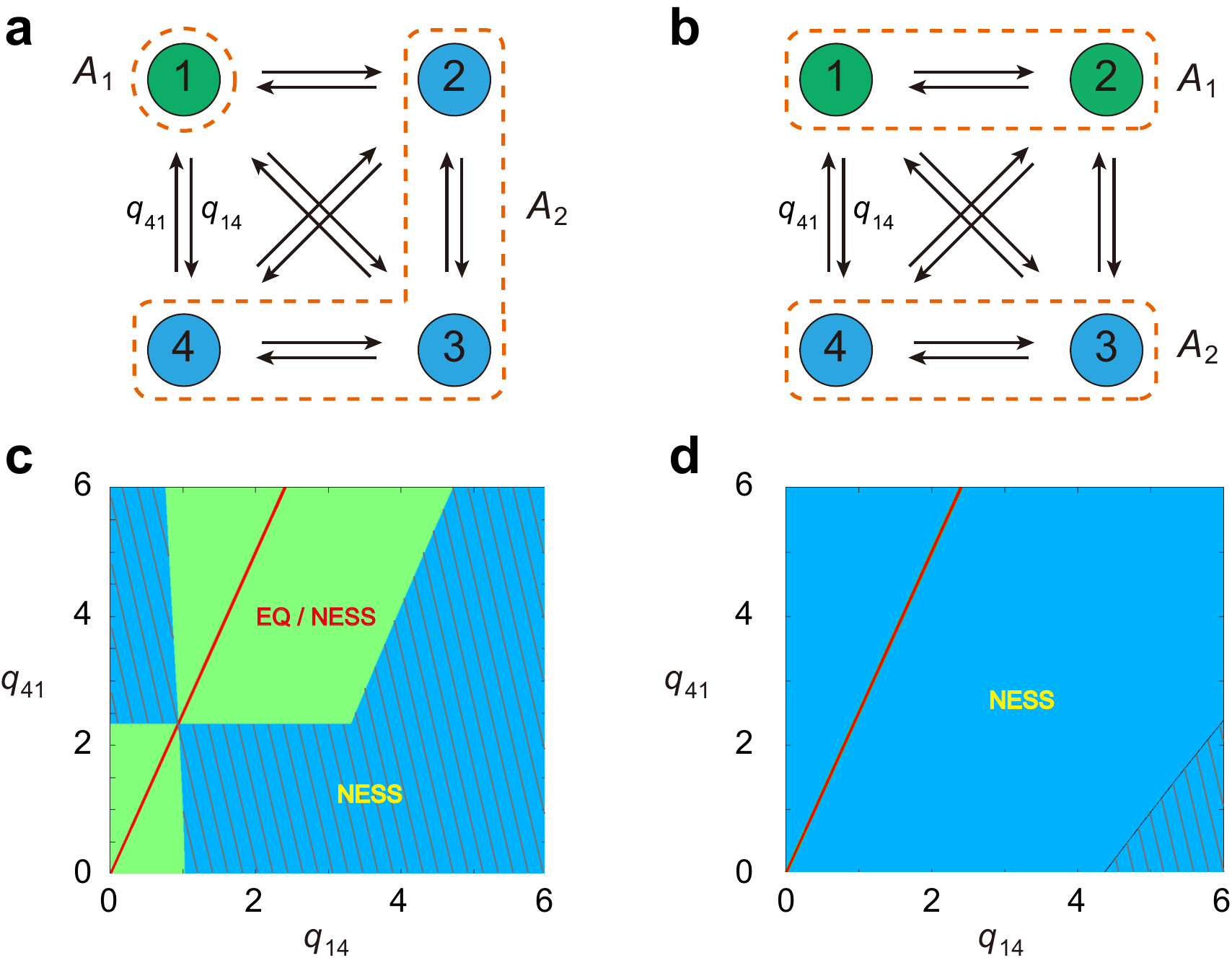}
\caption{\textbf{NESS identification for a four-state Markovian system.}
\textbf{(a)} System with coarse-grained partitioning $A_1=\{1\}$ and $A_2=\{2,3,4\}$.
\textbf{(b)} System with coarse-grained partitioning $A_1=\{1,2\}$ and $A_2=\{3,4\}$.
\textbf{(c)} Phase diagram in the $q_{14}-q_{41}$ plane for the system in (a).
\textbf{(d)} Same as (c) but for the system in (b). In (c),(d), the red line shows the region of equilibrium states (EQ). The blue (and shaded blue) area shows the NESS region that can be identified by our criterion. The shaded blue area shows the NESS region that can be identified by the autocorrelation criterion. The green area shows the NESS region that fails to be detected by our criterion. The parameters are chosen as $q_{12}=1$, $q_{13}=q_{21}=2$, $q_{23}=q_{24}=4$, $q_{31}=q_{32}=q_{34}=3$, and $q_{42}=q_{43}=5$.}\label{fig:heatmap}
\end{figure}

Finally, as an example, we focus on a four-state fully connected system with two different coarse-grained partitionings (Fig. \ref{fig:heatmap}(a),(b)). The system has $12$ unknown parameters and we want to determine whether it is in an NESS. We first consider the partitioning of $A_1=\{1\}$ and $A_2=\{2,3,4\}$ (Fig. \ref{fig:heatmap}(a)). In this case, our criterion is equivalent to the autocorrelation criterion. Fig. \ref{fig:heatmap}(c) illustrates the NESS region in the parameter space that can be identified by our criterion, or equivalently, the autocorrelation criterion (shown in shaded blue) and the region that fails to be identified by the two criteria (shown in green). The red line shows the region of equilibrium states. In this case, NESS can only be detected in the parameter region far from the red line.

We then consider another partitioning, i.e. $A_1=\{1,2\}$ and $A_2=\{3,4\}$ (Fig. \ref{fig:heatmap}(b)). Fig. \ref{fig:heatmap}(c) shows the NESS regions that can be identified by our criterion (shown in blue) and by the autocorrelation criterion (shown in shaded blue). Clearly, the entire NESS region can be captured by our criterion but only a small subregion can be detected by the autocorrelation criterion. To gain deeper insights, note that there are $S = 6$ sufficient statistics for the partitioning shown in Fig. \ref{fig:heatmap}(a), while for the one shown in Fig. \ref{fig:heatmap}(b), there are $S = 8$ sufficient statistics. More sufficient statistics means that more information can be extracted from coarse-grained observations, which generally leads to a larger NESS region that can be identified by our criterion.

\emph{Estimation of sufficient statistics} --- We emphasize that our methods of parameter inference and nonequilibrium detection are based on an accurate estimation of all the sufficient statistics. In the End Matter, we have proposed a maximum likelihood approach to accurately inferring the sufficient statistics $\Lambda_{A_iA_j}$ as well as the transition rates $q_{xy}$ (when $S\geq |E|$). This method is then validated using synthetic time-trace data for the ladder, cyclic, and linear models generated using stochastic simulations.

\emph{Conclusions and discussion} ---  Another crucial question beyond this study is whether the transition topology of the system can also be inferred based on coarse-grained observations. In fact, this is impossible in many cases due to the loss of information during coarse-graining; however, when there are two coarse-grained states, Refs. \cite{bruno2005using, flomenbom2005can, flomenbom2006utilizing, flomenbom2008universal} have developed a canonical form method of finding all possible transition topologies of the underlying Markovian dynamics. Hence in this paper, we always assume that the transition topology of the system is given.

Here we established a framework of parameter inference and nonequilibrium identification based on coarse-grained observations for a general Markovian system with an arbitrary number of microstates and an arbitrary number of coarse-grained states when the underlying transition topology is given. We provided a criterion for evaluating whether the coarse-grained information is enough for estimating all transition rates and also a criterion for detecting whether the system is in an NESS. Our method is based on extracting a set of sufficient statistics that incorporates all statistical information of an infinitely long coarse-grained trajectory. Using these sufficient statistics, the problems of parameter recovery and nonequilibrium detection can be brought into a unified theoretical framework.

\emph{Acknowledgements} --- C. J. acknowledges support from National Natural Science Foundation of China with grant No. U2230402 and No. 12271020.

\footnotesize\setlength{\bibsep}{5pt}
\bibliography{myref}
\bibliographystyle{apsrev4-2}

\normalsize
\section*{End matter}
\emph{Appendix A: Necessity of the NESS criterion} --- In the main text, we proposed an NESS criterion which states that the violation of the coarse-grained probability distribution condition (Eq. \eqref{fm:criterion1}) or the coarse-grained detailed balanced condition (Eq. \eqref{fm:criterion2}) implies the presence of nonequilibrium. Note that this is a sufficient condition for detecting nonequilibrium. Here we prove that for any three-state system with two coarse-grained states $A_1=\{1\}$ and $A_2=\{2,3\}$, the above criterion is also a necessary condition. To this end, we only need to show that if Eqs. \eqref{fm:criterion1} and \eqref{fm:criterion2} are both satisfied, then among all three-state systems having the same sufficient statistics $\tilde\Lambda_{A_iA_j}$, there must exist a system which is in equilibrium. In fact, if Eqs. \eqref{fm:criterion1} and \eqref{fm:criterion2} hold, then we prove in \cite{supp} that the entries of $\tilde{\Lambda}_{A_iA_j}$ are all positive. To proceed, we set
\begin{equation}
\Phi_{A_1} = 1,\;\;\;\Phi_{A_2} = \begin{pmatrix}
\frac{o_1\cos\theta}{\sqrt{\tilde{\alpha}_2^2}} & \frac{o_2\sin\theta}{\sqrt{\tilde{\alpha}_2^2}}\\
\frac{-o_1\sin\theta}{\sqrt{\tilde{\alpha}_3^2}} & \frac{o_2\cos\theta}{\sqrt{\tilde{\alpha}_3^2}}\\
\end{pmatrix},
\end{equation}
where $\theta$ is an undetermined constant and
\begin{equation}
\begin{split}
o_1 &= \sqrt{\tilde{\alpha}_2^2}\cos\theta-\sqrt{\tilde{\alpha}_3^2}\sin\theta,\\
o_2 &= \sqrt{\tilde{\alpha}_2^2}\sin\theta+\sqrt{\tilde{\alpha}_3^2}\cos\theta.
\end{split}
\end{equation}
Clearly, we have $\Phi_{A_2}\mathbf{1}^T = \mathbf{1}^T$. Moreover, we set
\begin{equation}\label{Qdef}
\begin{split}
Q = (q_{xy}) = \begin{pmatrix}
\Phi_{A_1}^{-1}\tilde{\Lambda}_{A_1}\Phi_{A_1} & \Phi_{A_1}^{-1}\tilde{\Lambda}_{A_1A_2}\Phi_{A_2} \\
\Phi_{A_2}^{-1}\tilde{\Lambda}_{A_2A_1}\Phi_{A_1} & \Phi_{A_2}^{-1}\tilde{\Lambda}_{A_2}\Phi_{A_2} \\
\end{pmatrix}.
\end{split}
\end{equation}
Direct computations show that
\begin{equation}\label{Qvalues}
\begin{split}
q_{12}&=\frac{o_1\cos\theta}{\sqrt{\tilde{\alpha}_2^2}}\tilde{\lambda}_{12}^{12}-\frac{o_1\sin\theta}{\sqrt{\tilde{\alpha}_3^2}}\tilde{\lambda}_{13}^{12},\\
q_{13}&=\frac{o_2\sin\theta}{\sqrt{\tilde{\alpha}_2^2}}\tilde{\lambda}_{12}^{12}+\frac{o_2\cos\theta}{\sqrt{\tilde{\alpha}_3^2}}\tilde{\lambda}_{13}^{12},\\
q_{21}&=\frac{\sqrt{\tilde{\alpha}_2^2}\cos\theta}{o_1}\tilde{\lambda}_{21}^{21}-\frac{\sqrt{\tilde{\alpha}_3^2}\sin\theta}{o_1}\tilde{\lambda}_{31}^{21},\\
q_{31}&=\frac{\sqrt{\tilde{\alpha}_2^2}\sin\theta}{o_2}\tilde{\lambda}_{21}^{21}+\frac{\sqrt{\tilde{\alpha}_3^2}\cos\theta}{o_2}\tilde{\lambda}_{31}^{21},\\
q_{23}&=\frac{o_2}{o_1}(\tilde{\lambda}^2_2-\tilde{\lambda}^2_1)\sin\theta\cos\theta,\\
q_{32}&=\frac{o_1}{o_2}(\tilde{\lambda}^2_2-\tilde{\lambda}^2_1)\sin\theta\cos\theta.
\end{split}
\end{equation}
Note that there are two cases: (i) if $\tilde{\lambda}_2^2>\tilde{\lambda}_1^2$, then we choose $\theta$ to be a very small positive number; (ii) if $\tilde{\lambda}_2^2<\tilde{\lambda}_1^2$, then we choose $\theta$ to be a very small negative number. For both the two cases, it is easy to check that $q_{xy}>0$ for any $x\neq y$ and hence $Q$ is indeed the generator matrix of a Markovian system. According to the definition of $Q$, it is clear that $\Lambda_{A_iA_j}(q_{xy}) = \tilde\Lambda_{A_iA_j}$. This shows that $\tilde\Lambda_{A_iA_j}$ are the sufficient statistics of the system.

On the other hand, it follows from Eq. \eqref{Qvalues} that
\begin{equation}\label{details}
\begin{split}
q_{12}q_{23}q_{31}
=&\;(\tilde{\lambda}^2_2-\tilde{\lambda}^2_1)\sin\theta\cos\theta\left(\frac{\cos\theta}{\sqrt{\tilde{\alpha}_2^2}}
\tilde{\lambda}_{12}^{12}-\frac{\sin\theta}{\sqrt{\tilde{\alpha}_3^2}}\tilde{\lambda}_{13}^{12}\right)\\
&\;\times \left(\sqrt{\tilde{\alpha}_2^2}\sin\theta\tilde{\lambda}_{21}^{21}
+\sqrt{\tilde{\alpha}_3^2}\cos\theta\tilde{\lambda}_{31}^{21}\right),\\
q_{13}q_{32}q_{21}
=&\;(\tilde{\lambda}^2_2-\tilde{\lambda}^2_1)\sin\theta\cos\theta\left(\frac{\sin\theta}{\sqrt{\tilde{\alpha}_2^2}}
\tilde{\lambda}_{12}^{12}+\frac{\cos\theta}{\sqrt{\tilde{\alpha}_3^2}}\tilde{\lambda}_{13}^{12}\right)\\
&\;\times \left(\sqrt{\tilde{\alpha}_2^2}\cos\theta\tilde{\lambda}_{21}^{21}
-\sqrt{\tilde{\alpha}_3^2}\sin\theta\tilde{\lambda}_{31}^{21}\right).
\end{split}
\end{equation}
Since the coarse-grained detailed balance condition holds, we have
\begin{equation}
\tilde{\alpha}^1_1\tilde{\lambda}^{12}_{12} = \tilde{\alpha}^2_2\tilde{\lambda}^{21}_{21},\;\;\;
\tilde{\alpha}^1_1\tilde{\lambda}^{12}_{13} = \tilde{\alpha}^2_3\tilde{\lambda}^{21}_{31}.
\end{equation}
This indicates that $\tilde{\alpha}^2_2\tilde{\lambda}^{21}_{21}\tilde{\lambda}^{12}_{13} = \tilde{\alpha}^2_3\tilde{\lambda}^{21}_{31}\tilde{\lambda}^{12}_{12}$. Inserting this into Eq. \eqref{details} yields $q_{12}q_{23}q_{31} = q_{13}q_{32}q_{31}$. This shows that the system satisfies the Kolmogorov cyclic condition and hence it is in equilibrium \cite{kolmogoroff1936theorie}.

\emph{Appendix B: Estimation of sufficient statistics} --- Our methods of parameter inference and nonequilibrium identification depend on an accurate estimation of all the sufficient statistics $\Lambda_{A_iA_j}$. Recall that once we have obtained estimates of $\Lambda_{A_iA_j}$, all the transition rates of the system can be determined by solving Eq. \eqref{fm:pa1}, and the system is in an NESS if any one of Eqs. \eqref{fm:criterion1} and \eqref{fm:criterion2} is violated. In applications, a crucial question is how to accurately estimate $\Lambda_{A_iA_j}$ using time-series measurements of coarse-grained states.

To answer this, here we assume that the coarse-grained states of the system can be observed at multiple discrete time points. The time resolution of the coarse-grained trajectory is assumed to be sufficiently high so that the jump times $0\leq t_1<...<t_m$ of the trajectory can be determined accurately (this is equivalent to saying that the waiting time distributions between coarse-grained states can be measured accurately, an assumption widely used in previous studies \cite{bruno2005using, flomenbom2005can, flomenbom2006utilizing, flomenbom2008universal}). The corresponding coarse-grained states before these jump times are denoted by $A_{i_1},\cdots,A_{i_m}$, respectively. For any Markovian system, we use the stochastic simulation algorithm to generate synthetic time-series recordings of coarse-grained states with exact $m$ jumps, where the number of jumps is chosen to be $m=3\times10^3,3\times10^4,3\times10^5$. In what follows, we refer to $m$ as the sample size.

We then use a maximum likelihood method to estimate these sufficient statistics. Recall that Eq. \eqref{likelihood} gives the probability density of each coarse-grained trajectory; hence for any given jump times $\mathcal{T} = \{t_1,\cdots,t_m\}$ and coarse-grained states $\mathcal{A}=\{A_{i_1},\cdots,A_{i_{m}}\}$, the likelihood function can be constructed as
\begin{equation}
\begin{split}
&\; L(\Lambda_{A_iA_j}|\mathcal{T},\mathcal{A})\\
& = \alpha_{A_{i_1}}e^{\Lambda_{A_{i_1}}t_1}\cdots\Lambda_{A_{i_{m-1}}A_{i_{m}}}e^{\Lambda_{A_{i_{m}}}(t_m-t_{m-1})}\mathbf{1}^T,
\end{split}
\end{equation}
where $\Lambda_{A_iA_j}$ are the sufficient statistics to be estimated, and we have shown that both $\alpha_{A_i}$ and $\Lambda_{A_i}$ are fully determined by $\Lambda_{A_iA_j}$. The estimates of the sufficient statistics can be obtained by maximizing the likelihood function, i.e.
\begin{equation}
\tilde\Lambda_{A_iA_j} = \underset{\Lambda_{A_iA_j}}{\mathrm{arg\;max}}\;L(\Lambda_{A_iA_j}|\mathcal{T},\mathcal{A}).
\end{equation}
The estimation accuracy of each entry $\lambda^{ij}_{kl}$ of $\Lambda_{A_iA_j}$ can be measured by the relative error
\begin{equation}
\kappa(\lambda^{ij}_{kl}) = \frac{\big|\tilde\lambda^{ij}_{kl}-\lambda^{ij}_{kl}\big|}{\big|\lambda^{ij}_{kl}\big|}.
\end{equation}
If the relative error is less than $0.2$, then we believe that the estimated value of $\lambda^{ij}_{kl}$ can reflect the realistic dynamic property of the system \cite{jiao2024can}. Moreover, let $\eta$ denote the proportion of those $\lambda^{ij}_{kl}$ ($1\leq i\neq j\leq n$, $k\in A_i$, $l\in A_j$) which have a relative error less than $0.2$. If $\eta\geq 75\%$, i.e. $75\%$ of sufficient statistics have a relative error less than $0.2$, then we believe that an accurate estimation of sufficient statistics is made.
\begin{figure}[htb!]
\centering\includegraphics[width = 0.48\textwidth]{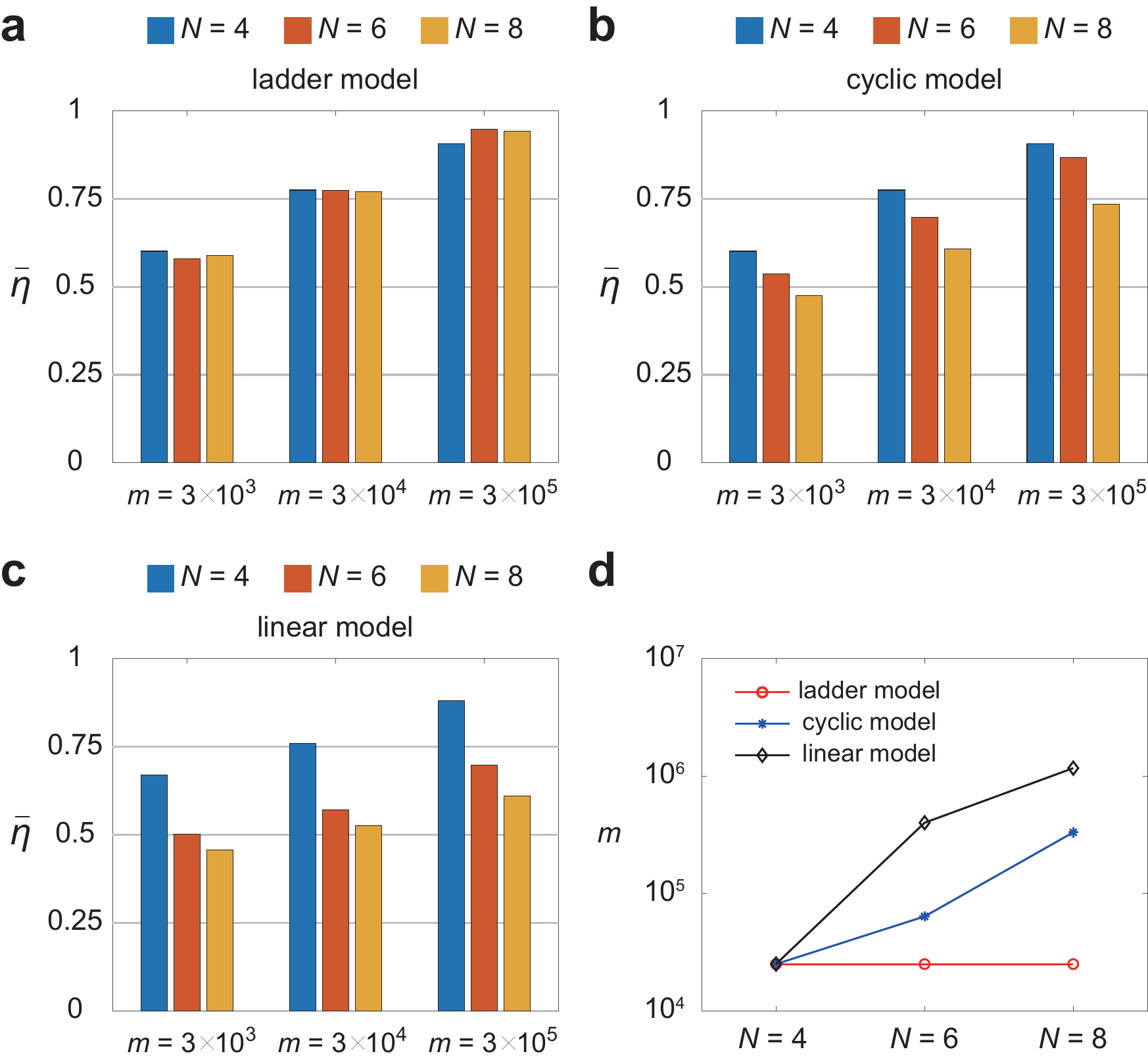}
\caption{\textbf{Accuracy of sufficient statistics estimation for the three models}.
\textbf{(a)} Plot of $\bar\eta$ as a function of $N$ and $m$ for the ladder model. Here $\bar\eta$, which characterizes the accuracy of estimation, is defined as the sample mean of $\eta$ for $100$ randomly selected parameter sets.
\textbf{(b)} Same as (a) but for the cyclic model.
\textbf{(c)} Same as (a) but for the linear model. In (a)-(c), all the transition rates $q_{xy}$ are randomly selected so that $\log_{10}q_{xy}\sim U[0,1]$, where $U[0,1]$ denotes the uniform distribution within the interval $[0,1]$.
\textbf{(d)} The minimal sample size $m$ required for achieving accurate estimation as a function of $N$ for the three models. Empirically, if $\bar\eta\geq 75\%$, then we believe that an accurate estimation is made for the corresponding model.}\label{estimate}
\end{figure}

Next we apply our inference method to three specific models: (i) the ladder model (Fig. \ref{model}(c)) with $N/2$ coarse-grained states
\begin{equation}
A_1=\{1,2\},\cdots,A_{N/2}=\{N-1,N\};
\end{equation}
(ii) the cyclic model (Fig. \ref{model}(d)) with two coarse-grained states
\begin{equation}
A_1=\{1,2\},\;\;\;A_2=\{3,\cdots,N\};
\end{equation}
(iii) the linear model (Fig. \ref{model}(f)) with two coarse-grained states
\begin{equation}
A_1=\{1\},\;\;\;A_2=\{2,3,\cdots,N\}.
\end{equation}
All these models satisfy $S\geq |E|$ and hence a complete parameter recovery is generally possible. For each model, we perform parameter inference using the maximum likelihood method under $100$ randomly selected parameter sets, which cover large swathes of parameter space.

Let $\bar\eta$ denote the sample mean of $\eta$ for all the $100$ parameter sets; clearly, it characterizes the estimation accuracy for the corresponding model. Fig. \ref{estimate}(a)-(c) show the value of $\bar\eta$ as a function of the number of microstates $N$ and the sample size $m$ for the three models. As expected, an increasing sample size leads to a higher estimation accuracy. Interestingly, we also find that the estimation accuracy, evaluated by $\bar\eta$, is insensitive to the number of microstates for the ladder model. while it is very sensitive to the number of microstates for the cyclic and linear models. This is possibly because the ladder model has more coarse-grained states $(n = N/2)$ than the cyclic and linear models ($n = 2$).

Another crucial question is what sample size is needed for an accurate estimation. Empirically, if $\bar\eta\geq 75\%$, then we believe that an accurate estimation is made for the corresponding model. Fig. \ref{estimate}(d) shows the minimal sample size $m$ required for achieving accurate estimation as a function of the number of microstates $N$. For the ladder model, the minimal sample size is roughly $m\approx 3\times 10^4$, insensitive to the number of microstates $N$. For the cyclic and linear models, the minimal sample size increases significantly with $N$. When $N = 4$, all the three models require a similar sample size of $m\approx 3\times 10^4$ for achieving accurate inference. However, when $N = 8$, the cyclic model requires a sample size of $m\approx 5\times 10^5$ and the linear model requires a sample size of $m\approx 10^6$.

Thus far, we only focus on the estimation of the sufficient statistics $\Lambda_{A_iA_j}$. Once the sufficient statistics have been determined, the transition rates $q_{xy}$ can also be recovered by solving Eq. \eqref{fm:pa1} numerically. Similarly, we can define the counterpart of $\bar\eta$ for transition rate estimation. Supplementary Fig. 1 shows the accuracy of transition rate estimation as a function of $N$ and $m$ for the three models. Comparing Supplementary Fig. 1 with Fig. \ref{estimate}, it can be seen that the accuracy of transition rate estimation is similar to but slightly lower than the accuracy of sufficient statistics estimation.

\end{document}